\documentclass[a4paper]{jpconf}
\usepackage{graphicx}
\begin{document}
\title{Constraints on the Equation-of-State of neutron stars from nearby neutron star observations}

\author{R Neuh\"auser, V V Hambaryan, M M Hohle, T Eisenbeiss}

\address{Astrophysikalisches Institut, University Jena,
Schillerg\"a\ss chen 2-3, D-07745 Jena, Germany}

\ead{rne@astro.uni-jena.de}

\begin{abstract}
We try to constrain the Equation-of-State (EoS)
of supra-nuclear-density matter in neutron stars (NSs)
by observations of nearby NSs.
There are seven thermally emitting NSs known from X-ray
and optical observations, the so-called Magnificent Seven (M7),
which are young (up to few Myrs), nearby (within a few hundred pc),
and radio-quiet with blackbody-like X-ray spectra, 
so that we can observe their surfaces.
As bright X-ray sources, we can determine their rotational
(pulse) period and their period derivative from X-ray timing.
From XMM and/or Chandra X-ray spectra, we can determine their temperature.
With precise astrometric observations using the Hubble Space
Telescope, we can determine their parallax (i.e. distance)
and optical flux. From flux, distance, and temperature, one
can derive the emitting area - with assumptions about the 
atmosphere and/or temperature distribution on the surface.
This was recently done by us for the two brightest M7 NSs RXJ1856 and RXJ0720. 
Then, from identifying absorption lines in X-ray spectra,
one can also try to determine gravitational redshift.
Also, from rotational phase-resolved spectroscopy,
we have for the first time determined the compactness (mass/radius) 
of the M7 NS RBS1223.
If also applied to RXJ1856, radius (from luminosity and temperature)
and compactness (from X-ray data) will yield the mass and radius -
for the first time for an isolated single neutron star.
We will present our observations and recent results.
\end{abstract}

\section{Introduction: The Magnificent Seven Neutron Stars}

With deep optical follow-up observations of 
bright X-ray sources from the ROSAT All-Sky Survey, seven neutron
stars (NSs) were identified, which share their main properties
(called {\em Magnificent Seven} M7),
but are different from radio pulsars:
Bright X-ray sources, faint optical sources, thermal X-ray spectra, X-ray pulsation,
isolated without companions nor nearby supernova remnants, no radio emission, 
strong magnetic fields, and intermediate spin-down ages.
See Haberl (2007) for a recent reviews.
The two brightest M7 NSs, RXJ1856 and RXJ0720, were identified as such
by Walter et al. (1996) and Haberl et al. (1997).

Mass determinations for M7 NSs would be of special interest, because
these M7 NSs may well be different from radio pulsars in binaries,
for which masses could be determined before from orbital solutions,
e.g. they do not show radio emission and rotate slower (few seconds)
than normal radio pulsars, but still have strong magnetic fields.

\section{Basic X-ray and optical observations to obtain the radius}

Since the X-ray spectra of the M7 NSs 
are consistent with blackbodies (some also show broad 
absorption features suggested to be proton cyclotron absorption), 
we observe their cooling surfaces.
With X-ray spectroscopy, we can determine the temperature of
the the X-ray emitting surface.
E.g., for RXJ1856 and RXJ0720, the blackbody temperatures
are $63 \pm 1$ eV for RXJ1856 (Burwitz et al. 2001, 2003)
and $90 \pm 1$ eV for RXJ0720 (Hohle et al. 2009).

The distance towards these two brightest M7 NSs
have been observed directly as trigonometric parallaxe
with the Hubble Space Telescope (HST),
even though they are faint (V=25.7 mag for RXJ1856
and V=26.8 mag for RXJ0720).
The results have been a matter of debate for the last years:
For RXJ1856, 
Kaplan et al. (2002) found $140 \pm 40$ pc with three HST PC data points;
Walter et al. (2002) got $117 \pm 12$ pc with four HST PC data points;
then after having obtained eight HST ACS data points,
Kaplan et al. (2007) obtained $167 \pm 17$ pc,
while Kaplan et al. (2007) gave $161 \pm 16$ pc as result;
the latter two papers,
however, did not give any details as to how the results were obtained;
then, Walter et al. (2010) re-reduced the eight HST ACS data points
with various techniques, all yielding $123 \pm 13$ pc.
The latter distance is also consistent with $135 \pm 25$ pc obtained
by Posselt et al. (2007) from a 3D N$_{\rm H}$ extinction model,
so that RXJ1856 is just foreground to the CrA dark cloud located
at around $\sim 130$ to $\sim 140$ pc.

For RXJ0720, Kaplan et al. (2007) published $\sim 360$ pc (250 - 530 pc)
based on eight HST ACS data points.
Posselt et al. (2007), using their
3D N$_{\rm H}$ extinction model, determined $\sim 250$ pc for the
distance towards RXJ0720.
Recently, Eisenbeiss (2011) re-determined the parallaxe of the NS
RXJ0720 using the archival eight HST data points yielding $\pi = 3.6 \pm 1.6$ mas
(Eisenbeiss 2011), i.e. $\sim 278$ pc (192-500 pc); this is consistent
with an earlier determination by Kaplan et al. (2007),
(270-530 pc) within large error bars. 

Once flux in different wavelength, extinction, and 
distance are known, one can calculate the lumonisity.
Luminosity and temperature can immediately
yield the emitting area if blackbody.
This would be true and correct, if both the X-ray emission
(used to determine the temperature) and the optical emission
(used to determine the distance) come from the same (emitting) area.
However, both RXJ1856 and RXJ0720, like all M7 NSs, show a 
so-called optical excess, i.e. they emit more optical emission
(by a factor of 7 for RXJ1856, Burwitz et al. 2003, 
and 10 for RXJ0720, Eisenbeiss et al. 2010)
than expected from extrapolating their X-ray blackbody spectra.
Hence, it might well be that the X-ray emission comes from
one or two small emitting areas, while the optical emission
comes from the whole warm surface.

Since the temperature distribution across the surface is not known,
e.g. the spot(s) could have some 
non-uniform temperature distribution with or without 
a thin atmosphere (e.g. H), there is so far no
unique way to derive the radius. 
It was determined first by Walter et al. (1999) obtaining $\sim 14$ km,
and most thoroughly by Tr\"umper (2003), who got $\sim 16.5$ km
(for radius at infinity, see Fig. 1)
from a two-blackbody optical and X-ray sepctral energy distribition at 117 pc,
and Ho et al. (2007), who found $\sim 17$ km ($\pm 30 \%$) for a 
magnetic H atmosphere (at 140 pc).
For any simple distribution, one would obtain
$16.8 \pm 1.25$ km radius for RXJ1856 (Walter et al. 2010),
using the most recent distance of $123 \pm 13$ pc, where the error bar includes
all errors from brighness, temperature (both around $1~\%$)
and distance ($\sim 10~\%$). 

\begin{figure}
\centering \includegraphics[width=9cm]{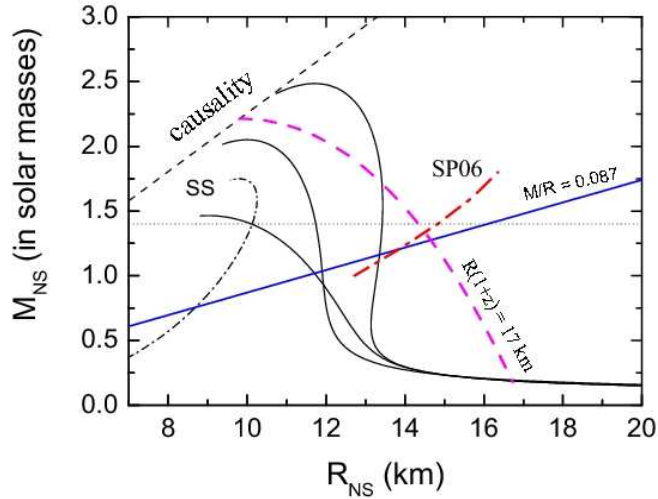}
\caption{\small
Mass-radius relations for NSs and several EoS (Haensel et al. 2007) as thin solid curves,
plus a strange star EoS as thin dash-dotted.
The thick dashed curve is for a constant radius 
at infinity of $R^\infty=R(1+z)=R/\sqrt{1-2GM/Rc^2}=17 \pm 3$ km
as found for both RXJ1856 and RXJ0720, see Sect. 2.
The thick dash-dotted line gives a model-dependent estimation of the M-R relation
from boundary layer spectra modeling in LMXBs (Suleimanov \& Poutanen 2006).
The straight solid line gives M~[M$_{\odot}$]~/~R~[km]=0.087
from rotational phase-resolved X-ray spectroscopy of the M7 NS RBS 1223.}
\end{figure}

For RXJ0720, one obtains a radius (at infinity) of $17 \pm 3$ km (280 pc distance), 
where the error budget does not yet include the large distance error.

Those two radius determinations (both radii as seen at infinity, see Fig. 1)
are so far the only two direct radius measurements for NSs.
Given that the peaks in the optical emission are not seen in both NSs,
those estimates might be considered as lower radius limit.

\section{X-ray spectroscopy to obtain the redshift}

Both RXJ1856 and RXJ0720, as well as other M7 NSs, are observed often with
both the X-ray satellites XMM and Chandra, partly for (cross-)calibration purposes. 
Hence, one can combine all spectroscopic observations of either XMM or Chandra
to obtain a spectrum with high resolution and high S/N.
A measurement of the gravitational redshift of spectral features in the spectrum of thermal emitting NSs
would yield the so-called compactness, i.e. mass-to-radius ratio.

Hambaryan et al. (2009) detected a narrow absorption feature at 0.57 keV in the co-added XMM/RGS spectrum of RXJ0720.
The feature was identified with an absorption line of highly ionized oxygen O VII resonance K$\alpha$ line, possibly originating in 
the interstellar medium (in the foreground of RXJ0720) or even more likely in the nearby ambient circumstellar medium around RXJ0720.
If the line would have originated in the thin NS atmosphere, it would be very difficult to
identify it, because the gravitational redshift is still poorly known for very large magnetic fields
like in M7 NSs. A redshift of $\sim 0.16$ (Hambaryan et al. 2009) would 
correspond to a compactness of mass (in solar masses)
devided by radius (in km) of $\sim 0.087$ (e.g., for the canonical NS mass of 1.4 solar masses, this
would correspond to 16 km radius). In the latter case, one should also detect other lines 
redshifted by the same amount, but they are not detected.

\section{X-ray rotational phase resolved spectroscopy to measure the compactness}

The light curve of the M7 NS RBS1223 shows not only the highest pulse fraction with $18~\%$ 
(Schwope et al. 2007),
but also different double-humped light curves at different energy intervalls. This behaviour may
indicate the existence of two emitting areas, which are not exactly identical in size, temperarure, and 
location with respect to the rotation axes, e.g. they could be non-antipodal
with a small shift angle. Hence, it should be possible to perform rotational phase resolved
spectroscopy, i.e. to investigate (and fit) light curves at different energy intervalls and
spectra at different spectral ranges simultaneously. 
Given that RBS1223 also shows the highest pulse fraction, it is best suited for such an attempt.

Hambaryan et al. (2011) could determine from the best fit several parameters
of the NS: Not only the sizes, temperatures, magnetic fields, and locations of the two emitting areas,
together with the energy and strength of the broad proton cyclotron absorption feature
and the absorbing foreground column density, but also 
the gravitational redshift or compactness of the NS.

The result here is gravitational redshift $z = 0.16 \pm 0.02$ 
(or $0.15 \pm 0.02$ for a condensed iron surface with partially ionized H atmosphere
as described in Suleimanov et al. (2010), or $0.17 \pm 0.03$ for a pure blackbody
with electron scattering) corresponding to a
compactness (mass/radius) of $\sim 0.087 \pm 0.011$, e.g. a radius of $16 \pm 1$ km for 1.4 solar masses.

\section{Summary and Outlook}

The first two radius determinations for NSs, namely for RXJ1856 and RXJ0720, provide 
strong constraints for the EoS, as they exclude quark stars, but are consistent
with a very stiff EoS. Even if these radii are interpreted as lower limits,
due to the uncertainties in the model assumptions,
they still exclude quark or strange stars with Pion and Kaon condensates.

The attempts to identify atomic absorption lines in M7 X-ray spectra to
determine the gravitational redshift, have not yielded unique results so far,
because the detected faint lines could also have originated in the
interstellar or circumstellar material.

Rotational phase resolved spectra yield the gravitational redshift
or compactness of M7 NSs.
Combined with a radius (or mass) determination,
such as the ones above for RXJ1856 or RXJ0720,
they would provide the possibility to determine for the very same
NS(s) at the same time both radius and mass.
This would be a stringent constraint.

\subsection{Acknowledgments}
We would like to thank DFG in SFB TR 7 for financial support.


\section{References}

\begin{thebibliography}{}

\bibitem{} Burwitz V, Zavlin V E, Neuh\"auser R, et al. 2001 {\it A\&A} {\bf 379} L35

\bibitem{} Burwitz V, Haberl F, Neuh\"auser R, et al. 2003 {\it A\&A} {\bf 399} 1109

\bibitem{} Eisenbeiss E 2011 PhD thesis U Jena

\bibitem{} Eisenbeiss T, Ginski C, Hohle M M, Hambaryan V V, Neuh\"auser R, Schmidt T O B 2010 AN {\it 331} 243

\bibitem{} Haberl F 2007 {\it ApSS} {\bf 308} 181

\bibitem{} Haberl F, Motch C, Buckley D A H, Zickgraf F J, Pietsch W. 1997 {\it A\&A} {\bf 326} 662

\bibitem{} Haensel P, Potekhin A Y, Yakovlev D G 2007 {\it ASSL} {\bf 326} 

\bibitem{} Hambaryan V V, Neuh\"auser R, Haberl F, Hohle M M, Schwope A 2009 {\it A\&A} {\bf 497} L9

\bibitem{} Hambaryan V V, Suleimanov V, Schwope A P, Neuh\"auser R et al. 2011 {\it A\&A} {\bf 534}, 74

\bibitem{} Ho W C G, Kaplan D L, Chang P, van Adelsberg M, Potekhin A Y 2007 {\it MNRAS} {\bf 375} 821

\bibitem{} Hohle M M, Haberl F, Turolla R, Hambaryan V V, Zane Sm de Vries C P, Mendez M 2009 {\it A\&A} {\bf 498} 811

\bibitem{} Kaplan D L, van Kerkwijk M H, Anderson J 2002 {\it ApJ} {\bf 571} 447

\bibitem{} Kaplan D L, van Kerkwijk M H, Anderson J 2007 {\it ApJ} {\bf 660} 1428


\bibitem{} Posselt B, Popov S B, Haberl F, et al. 2007 {\it ApSS} {\bf 308} 171

\bibitem{} Schope A, Hambaryan V V, Haberl F, Motch C 2007 {\it ApSS} {\bf 308} 619

\bibitem{} Suleimanov V \& Poutanen 2006 {\it MNRAS} {\it 369} 2036

\bibitem{} Suleimanov V, Hambaryan V V, Potekhin A Y, van Adelsberg M, Neuh\"auser R, Werner K 2010 {\it A\&A} {\bf 522} 111

\bibitem{} Tr\"umper J, Burwitz V, Haberl F, et al. 2003 {\it Nucl. Phys. B Proc. Suppl.} {\bf 132} 560

\bibitem{} Walter F W, Wolk S J, Neuh\"auser R 1996 {\it Nature} {\bf 379} 233

\bibitem{} Walter F W \& Matthews L D 1999 {\it Nature} {\bf 389} 358

\bibitem{} Walter F W \& Lattimer J M 2002 {\it ApJ} {\bf 576} L145

\bibitem{} Walter F W, Eisenbeiss T, Lattimer J M, et al. 2010 {\it ApJ} {\bf 724} 669

\end{thebibliography}
\end{document}